# Stress-Dependent Magnetoimpedance in Co-Based Amorphous Wires and Application to Tunable Microwave Composites

Serghei Sandacci, Dmitriy Makhnovskiy, Larissa Panina, and Vladimir Larin

*Abstract* **A remarkably strong dependence of magnetoimpedance (MI) on tensile stress has been observed in the microwave frequency range for thin CoMnSiB glass-coated microwires exposed to a special thermal treatment. The MI ratio runs into more than 100% at 0.5–1.5 GHz when the tensile stress of 600 MPa is applied to the wire. It was demonstrated that a large MI change at such high frequencies is related predominantly with the dc magnetization orientation. A host of such microwires incorporated into a dielectric matrix may constitute a new sensing medium that is characterized by the stress-dependent effective permittivity. Such medium can be used for the microwave visualization of the stress distribution inside of a composite structure or on its surface.**

The magnetoimpedance (MI) effect in fine amorphous wires has been extensively investigated with relation to its application in various stress sensors.[1,2] An additional aspect of MI in wires that is of current research interest arises out of unique electromagnetic properties of wire-based composite materials. The use of ferromagnetic wires as filling elements in a dielectric matrix $\varepsilon$ makes it possible to achieve a large adjustability of the effective electromagnetic response.[3-6] These composites might have a particular interest for remote nondestructive testing, when the stress distribution inside the composite structure or on its surface can be visualized in the microwave range. In this work, the composite structure with short wires has been considered, which demonstrates the Lorentz dipole dispersion of $\varepsilon_{eff}$.[4] Short wire inclusions play a role of elementary dipole scatterers, which are polarized with an ac electric field. The collective microwave dipole response from the composite in free space can be characterized by $\varepsilon_{eff}$ having a resonant or relaxation dispersion depending on the current distribution in the wire and its impedance. At certain conditions, the impedance of ferromagnetic wires is very sensitive to the external factors, such as, the dc magnetic field or tensile stress. In the vicinity of the antenna resonance $f_{res} = c/(2l\sqrt{\varepsilon})$ even small variations in the wire impedance will result in a considerable change in the current distribution along the wire and, consequently, in the induced dipole moment of an elementary wire-scatterer. This effect has been proposed to be used as a stress-sensing mechanism in the composite structures.[5,6]

In this paper, we have investigated the MI effect in CoMnSiB glass-coated amorphous wires with thermally induced near-axial anisotropy. This wire shows a dc magnetization process which is dependent on a tensile stress, changing from almost a rectangular loop without a stress to a linear one when a strong stress is applied. This behavior in dc magnetization is important to obtain high sensitivity of microwave wire impedance to the applied tensile stress. The normalized impedance change is in the range of 100% for a characteristic stress of 600 MPa at 0.5-1.5 GHz. This result is in a good agreement with the theory. The stress-impedance data have been then used to model the effective dielectric response from composite containing short wire inclusions. We have demonstrated that the losses determined by the imaginary part of the effective permittivity can change more than twice as a result of applied tensile stress.

In general, the magnetoimpedance depends on both the static magnetization angle $\theta$ and the dynamic permeability $\mu_{eff}$, which has a very broad dispersion region. However, the theoretical and experimental analysis conducted in [4-7] has shown that $\mu_{eff}$ changes very little with both the external axial magnetic field $H_{ex}$ and anisotropy field $H_K$ for frequencies much higher than the ferromagnetic resonance, although preserving its relatively high values (in the range of tens). Therefore, the dependence of the magnetoimpedance on various external factors such as a dc magnetic field or stress (which changes $H_K$) will be entirely determined by the static magnetization angle $\theta$. In this case, provided that the skin effect is strong, the magnetoimpedance is proportional to $\cos^2(\theta)$. Such a functional dependence is referred to as "valve-like" behavior of MI at microwaves. Previously, this effect has been also observed experimentally in [8,9] for Co-

D. Makhnovskiy and L. Panina are with the School of Computing, Communications and Electronics, University of Plymouth, Plymouth, PL4 8AA, UK. (e-mails: dmakhnovskiy@plymouth.ac.uk, lpanina@plymouth.ac.uk). S. Sandacci is with Sensor Technology Ltd, Banbury, Oxon, OX15 6JB, UK. (e-mail: ss@sensors.co.uk). V. Larin is with MFTI Ltd, 26 Gurie Grosu, 39th Apartament, 2021 Kishinev, Republic of Moldova. (e-mail: mfti@company.md)

rich amorphous wires, but without a proper understanding of its origin.

In amorphous ferromagnetic alloys, the preferable magnetization direction is set by the combined effect of the shape anisotropy and the magnetoelastic anisotropy arising from the coupling between magnetostriction $\lambda$ and internal stresses $\sigma$ (axial, radial, and azimuth). The stresses are induced during the fabrication process and can be controlled by further post production treatment like drawing and annealing. Typically, the experimental hysteresis loops are nearly rectangular for $\lambda > 0$ and nearly linear for $\lambda < 0$. Because of the wire geometry, the stress transferred from the composite matrix to the wire inclusions is expected to be predominantly of a tensile nature. As it was noted above, to realize large impedance change, the applied tension needs to cause a directional change in the equilibrium magnetization. In negative magnetostrictive wires ($\lambda < 0$) with nearly circumferential anisotropy this can be achieved in the presence of a dc axial magnetic field, as it has been verified by measurements in [5]. To avoid the use of the dc bias field, which would be important for many practical applications, a special anisotropy, namely nearly axial for $\lambda < 0$, is required. Such a "reverse" anisotropy seems to be quite unusual; however, as it was reported recently in [10] on the example of Fe-based wires, it could be established by special annealing treatments. In this work, a nearly axial anisotropy in CoMnSiB glass-coated wires with the metallic core diameter $d$ of 10 microns was induced during annealing of a wire skein mounted on the aluminium bobbin (170$^o$C for 20 min). This was confirmed by hysteresis loop measurements.

For the analysis of the dc magnetization process we assume that there is an axial induced anisotropy with the anisotropy constant $K$, the axial stress $\sigma_a = \sigma_{res} + \sigma_{ex}$ which includes the residual $\sigma_{res}$ and external $\sigma_{ex}$ components, and also the torsion. The latter introduces a hysteresis in the model and can be decomposed into two perpendicular stresses $\pm\sigma_t$, each acting at $45°$ relative to the longitudinal wire axis. The positive $\sigma_t$ is the tensile stress, whereas the negative $-\sigma_t$ is the compressive stress. For this model, the magneto-static energy $U_{st}$ has the following form:

$$U_{st} = -K\cos^2\theta - \frac{3}{2}\lambda\sigma_a\cos^2\theta - \\ -\frac{3}{2}\lambda\sigma_t\left(\cos^2(\theta - 45°) - \cos^2(\theta + 45°)\right) - M_0 H_{ex}\cos(\theta) \quad (1)$$

where $\theta$ is the angle between the equilibrium magnetization $M_0$ and the wire axis, and $-M_0 H_{ex}\cos(\theta)$ is the Zeeman energy. After some algebra, Eq. (1) can be converted into the equivalent uniaxial anisotropy form with the anisotropy angle $\alpha$ that also is laid from the wire axis:

$$U_{st} = -\left|\tilde{K}\right|\cos^2(\alpha - \theta) - M_0 H_{ex}\cos(\theta)\,. \quad (2)$$

Here:

$$\tilde{K} = \frac{K + (3/2)\lambda\sigma_a}{\cos(2\tilde{\alpha})}, \quad \tilde{\alpha} = \frac{1}{2}\tan^{-1}\frac{3\left|\lambda\sigma_t\right|}{\left|K + (3/2)\lambda\sigma_a\right|}, \quad (3)$$

and the anisotropy angle is chosen as following:

$a)\ K + (3/2)\lambda\sigma_a > 0, \quad \alpha = \tilde{\alpha}$,

$b)\ K + (3/2)\lambda\sigma_a = 0, \quad \alpha = 45°, \quad \tilde{K} = 3\lambda\sigma_t$,

$c)\ K + (3/2)\lambda\sigma_a < 0, \quad \alpha = 90° - \tilde{\alpha}$.

For the negative magnetostriction, if $K - (3/2)|\lambda|\sigma_a > 0$ the anisotropy axis has an angle $\alpha < 45°$ with the wire axis. Then, with increasing the axial tension the anisotropy axis will rotate towards the circumferential direction. This corresponds to the behavior of the dc magnetization loops measured under the effect of stress.

Figure 1 shows the hysteresis curves of the annealed wire samples loaded at the end with a weight (1–8 g) creating a tensile stress of up to 700 MPa. In the case of stress-free wire, the hysteresis loop is nearly rectangular with the remanence-to-saturation ratio of about 0.8, which confirms that the wire has a nearly axial anisotropy. Application of the axial tension gradually turns the anisotropy back to the circumference and the loop is nearly linear (remanence-to-saturation drops down to 0.15). This process could be well described within the model introduced above that assumes the existence of a uniaxial anisotropy (of any origin as due to frozen-in compressive stresses, creep anisotropy or surface crystallization) and the anisotropy due to magnetostrictive interaction with applied tension and residual torsion. In the case of negative magnetostriction, the effective easy axis will gradually change from axial direction to the circumferential one by applying a tensile stress.

Figure 2 shows plots of high-frequency impedance vs. stress for frequencies 0.5, 1 and 1.5 GHz. The complex-valued impedance is found by measuring $S_{11}$ parameter with a specially designed microstripe cell [5] to minimize the post calibration mismatches originated by bonding the sample.

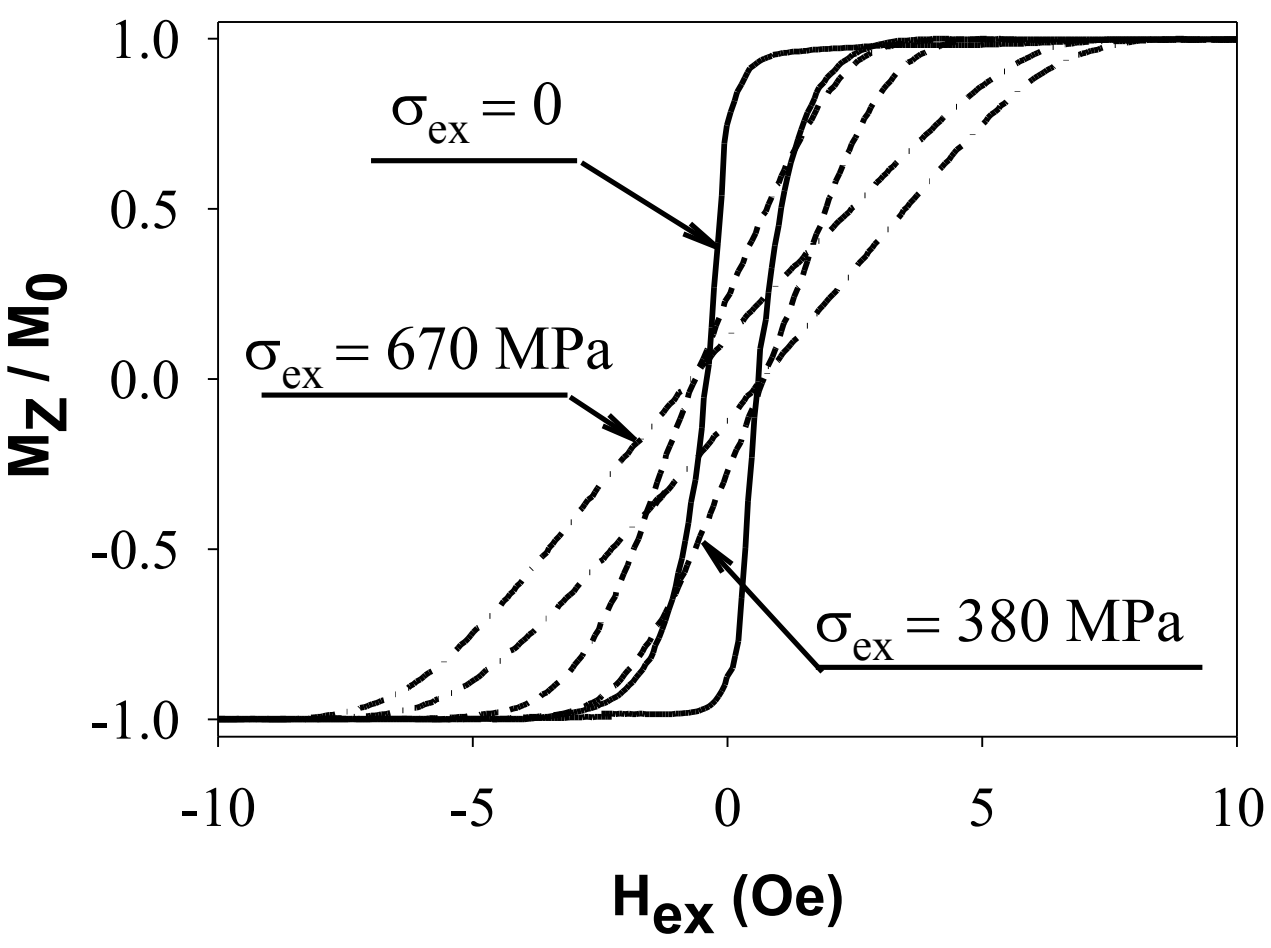

Fig. 1. Hysteresis curves of annealed CoMnSiB wires with the applied tension $\sigma_{ex}$ as a parameter.

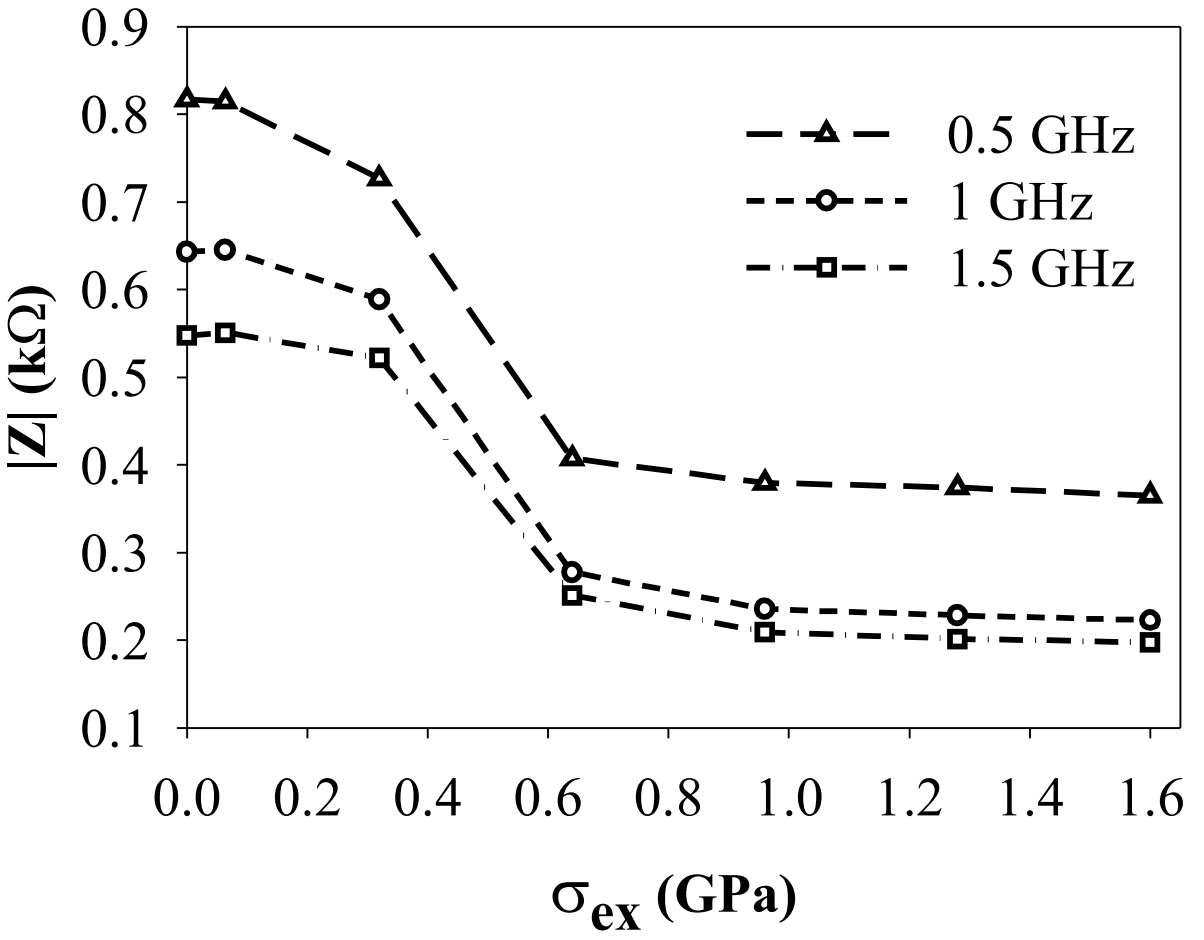

Fig. 2. Impedance plots vs. $\sigma_{ex}$ for different frequencies.

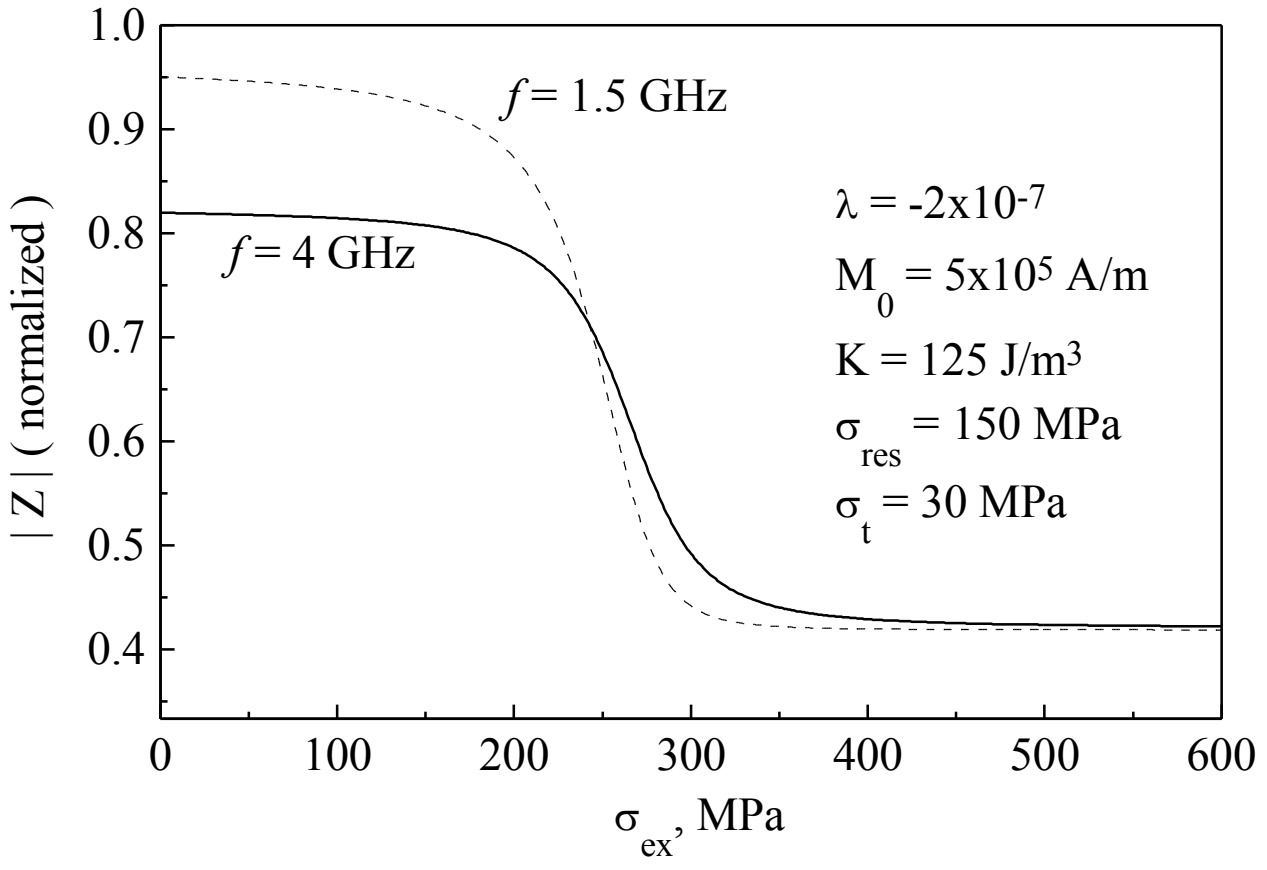

Fig. 3. Results of the magnetoimpedance vs. stress modeling.

It is seen that the wire impedance drops almost twice for 600 MPa at $0.5-1.5$ GHz, as a result of dc loop transformation. For higher stresses when the anisotropy becomes nearly circumferential the change in the impedance is not noticeable. The experimental stress dependences in Fig. 2 are well described by the standard model of magnetoimpedance [11] in conjunction with the magneto-static energy in the form of Eq. (2), as demonstrated in Fig. 3.

Figures 4(a,b) show the results of modeling of $\varepsilon_{eff} = \varepsilon^{/}_{eff} + i\varepsilon^{//}_{eff}$ in composite materials containing short pieces of ferromagnetic wires. The formula for $\varepsilon_{eff}$ was derived in [5] for the case of a very low wire concentration. Taking the wire length $l$ of 2 cm and the permittivity of the dielectric host material $\varepsilon$ of 4, the dipole resonance in wires which is responsible for the dispersion of $\varepsilon_{eff}$ is located in the vicinity of 4 GHz. The wire volume concentration $p$ was chosen 0.001% that is much below than the percolation threshold $p_c \propto d/l$. It is seen that the application of the external tensile stress $\sigma_{ex}$ greatly changes the dispersive behavior of $\varepsilon_{eff}$.

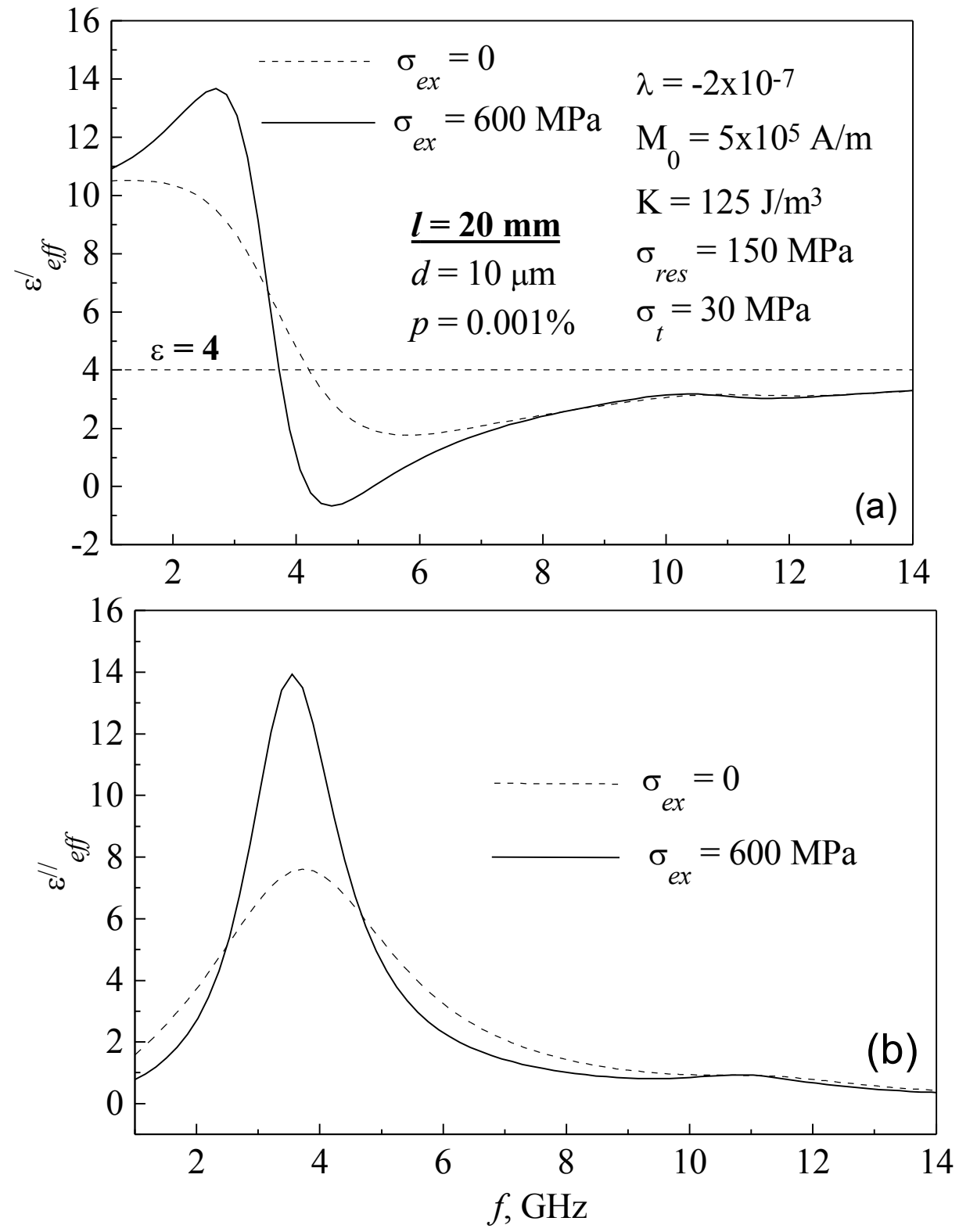

Fig. 4. The modeled dispersion of the effective permittivity with the external stress as a parameter.